\documentclass[aps,prc,twocolumn,superscriptaddress]{revtex4-2}
\usepackage{graphicx}% Include figure files
\usepackage{makecell}
\usepackage{color}
\usepackage[dvipsnames]{xcolor}
\usepackage[colorlinks=true,urlcolor=Blue,linkcolor=Blue,citecolor=Blue]{hyperref}
\usepackage[all]{hypcap}
\usepackage{amsmath}
\begin{document}

%\begin{frontmatter}

%\title{Effective stellar lifetime of $^{68}{\rm Se}$: a potential waiting-point in $rp$-process }

\title{Impact of experimental mass of $^{70}$Kr on the $^{68}$Se waiting-point in $rp$-process }

\author{M.~Zhang}
\affiliation{State Key Laboratory of Heavy Ion Science and Technology, Institute of Modern Physics, Chinese Academy of Sciences, Lanzhou 730000, China}
% \affiliation{School of Nuclear Science and Technology, University of Chinese Academy of Sciences, Beijing 100049, China}

\author{Y.~Luo}
\email{yudong.luo@pku.edu.cn}
\affiliation{School of Physics, Peking University, Beijing 100871, China}
\affiliation{Kavli Institute for Astronomy and Astrophysics, Peking University, Beijing 100871, China}

\author{A.~Dohi}
\email{akira.dohi@riken.jp}
\affiliation{RIKEN Pioneering Research Institute (PRI), 2-1 Hirosawa, Wako, Saitama 351-0198, Japan}
\affiliation{RIKEN Center for Interdisciplinary Theoretical \& Mathematical Sciences (iTHEMS), RIKEN 2-1 Hirosawa, Wako, Saitama 351-0198, Japan}

\author{X.~Xu}
\email{xuxing@impcas.ac.cn}
\affiliation{State Key Laboratory of Heavy Ion Science and Technology, Institute of Modern Physics, Chinese Academy of Sciences, Lanzhou 730000, China}
\affiliation{School of Nuclear Science and Technology, University of Chinese Academy of Sciences, Beijing 100049, China}

\author{X.L.~Yan}
\affiliation{State Key Laboratory of Heavy Ion Science and Technology, Institute of Modern Physics, Chinese Academy of Sciences, Lanzhou 730000, China}
\affiliation{School of Nuclear Science and Technology, University of Chinese Academy of Sciences, Beijing 100049, China}

\author{T.~Kajino}
\affiliation{School of Physics, Peng Huanwu Collaborative Center for Research and Education, and International Research Center for Big-Bang Cosmology and Element Genesis, Beihang University, Beijing 100191, China}
\affiliation{Graduate School of Science, The University of Tokyo, 7-3-1 Hongo, Bunkyo-ku, Tokyo 113-033, Japan}
\affiliation{Division of Science, National Astronomical Observatory of Japan, 2-21-1 Osawa, Mitaka, Tokyo 181-8588, Japan}

\author{Y.~H.~Zhang}
\affiliation{State Key Laboratory of Heavy Ion Science and Technology, Institute of Modern Physics, Chinese Academy of Sciences, Lanzhou 730000, China}
\affiliation{School of Nuclear Science and Technology, University of Chinese Academy of Sciences, Beijing 100049, China}
\author{M.~Wang}
\affiliation{State Key Laboratory of Heavy Ion Science and Technology, Institute of Modern Physics, Chinese Academy of Sciences, Lanzhou 730000, China}
\affiliation{School of Nuclear Science and Technology, University of Chinese Academy of Sciences, Beijing 100049, China}

\date{\today}

\begin{abstract}
The recent mass measurement of $^{70}$Kr using the $B\rho$-defined isochronous mass spectrometry yields a  mass excess of $-41320(140)$ keV, indicating a 220-keV increase in binding energy compared to the AME2020 prediction. 
%With this experimental mass, we re-evaluated the effective stellar half-life of
We utilize this experimental mass---the last piece of information needed---to model the potential waiting point $^{68}$Se in $rp$-process and quantitatively constrain the sequential $p$-capture reaction flow bypassing this waiting point. 
Our investigation shows that the more tightly bound nature of $^{70}$Kr enhances this reaction flow up to a factor of four. This enhancement reduces the effective half-life of $^{68}$Se.
{A} one-zone X-ray burst model calculations reveal that the higher flow of $^{70}$Kr has distinct effects on the tail structure of light curve and the final SnSbTe abundances in the ashes due to a stronger $rp$-process heating.
%The dominate uncertainty in determining the effective half-life of and reaction flow around $^{68}$Se originates from the large experimental error in the $^{70}$Kr mass. A more precise $^{70}$Kr mass measurement is highly desired.
\end{abstract}

%\begin{keyword}
%{Heavy-ion storage ring \sep Isochronous Mass Spectrometry }
%\end{keyword}

%\end{frontmatter}

\maketitle

\section{Introduction}\label{sec:introduction}

Type-I x-ray bursts (XRBs) are thermonuclear explosions occurring on the surface of a neutron star accreting hydrogen-rich matters from a companion star~\cite{Woosley1976,Joss1977,Fujimoto1981,Lewin1993,Parikh2013}. %As
The accreted material accumulates on the envelope of the neutron star, which leads to the high temperature and density conditions as $T\approx0.2~{\rm GK}$ and $\rho\approx10^6~{\rm g~cm^{-3}}$, respectively. Then, explosive nuclear burning is triggered by the triple-$\alpha$-reaction and CNO burning. CNO cycle could help in reaching the ignition temperature of XRB, and after its breakout around $T\approx0.5~{\rm GK}$, the nucleosynthesis moves to heavy proton-rich elements due to abundantly available protons, which
%the nuclear burning initiates through  $pp$-chains, the triple-$\alpha$-reaction, and the CNO cycles. The high temperatures and densities, created by released energy of the burning, are sufficiently hot to synthesize the heavier elements 
%via rapid fusion of abundantly available protons and seed nuclei formed from initial burning. 
%These nucleosynthesis processes 
are mainly the $\alpha p$-process (a sequence of proton captures and ($\alpha$, p) reactions) and the rapid proton capture process (a sequence of proton captures and subsequent $\beta^{+}$-decays, hereafter, $rp$-process)~\cite{Wallace1981,Wormer1994,SCHATZ1998167,Woosley2004,Fisker2008}. 
The amount of nuclear energy release can be directly observed as an XRB light curve such as the burst duration of $\sim\mathrm{a~few~} 10$ s \cite{Parikh2013,2025PASJ..tmp...39T}.
%and the time scale for the thermal runaway ranges between 10 seconds and several minutes~\cite{Parikh2013}. 
In order to quantitatively understand the shape of the burst light curve and the composition of the neutron star crust, detailed nuclear reaction network simulations are required. 
Of particular importance to the XRB models are the so-called waiting points in the $rp$-process~\cite{SCHATZ1998167}, which are essential to qualitatively explain the extended tails of observed light curves such as the well-known clocked burster GS 1826$-$24 (e.g., \cite{2007ApJ...671L.141H,2018ApJ...860..147M,Dohi2021}).

{A waiting point is characterized by low or even negative proton-capture reaction $Q$ values and relatively long $\beta$-decay half-life. 
During the $r$p-process nucleosynthesis, a thermodynamic equilibrium between forward proton capture and reverse photodisintegration reactions is established at the waiting point. This equilibrium induces process stagnation until $\beta$-decay of the waiting-point nuclide resumes the reaction chain. Therefore, the reaction flow towards heavier elements will be hindered by delaying the burning at the waiting points, leading to a often observed long burst tail in light curve.}

{The effective half-life, which is determined by the rates of $\beta$-decay and proton capture process, qualifies duration that takes to proceed a waiting point. As a result, it determines the extent to which the waiting point alter the observable x-ray light curve and the final abundance distribution of the burst. Notably, a sequential $p$-capture reactions can dramatically reduce effective lifetimes of waiting points (we call this process as ``2$p$-capture process'' for simplicity throughout this article). For nuclei with marginal binding energies, photodisintegration-driven equilibrium between ($Z,N$) and ($Z+1,N$) isotone neighbours enables sequential proton capture via the ($Z+1,N$) species. This bypass mechanism accelerates process progression despite the presence of classical waiting points. Reaction rates of both proton capture and 2$p$ capture in $r$p-process are exponentially dependent on the proton separation energies of the nuclides involved~\cite{SCHATZ1998167,Clark2004}, 
making precise nuclear mass measurements essential for accurate modeling of XRB nucleosynthesis. In recent years, various mass measurements with accuracy of better than 10-100 keV for neutron-deficient isotopes around three long-lived waiting points, $^{64}$Ge, $^{68}$Se, and $^{72}$Kr, have been established to explain the shape and duration of the observable x-ray light curve of the burst~\cite{Clark2004,Savory2009,Schury2007,Rodriguez2004,Tu2011,Cyburt2016,Schatz2017,Zhou2023}.}
%\cite{Schatz2001}
\begin{figure}[!htb]
	\centering
	\includegraphics[width=0.46\textwidth]{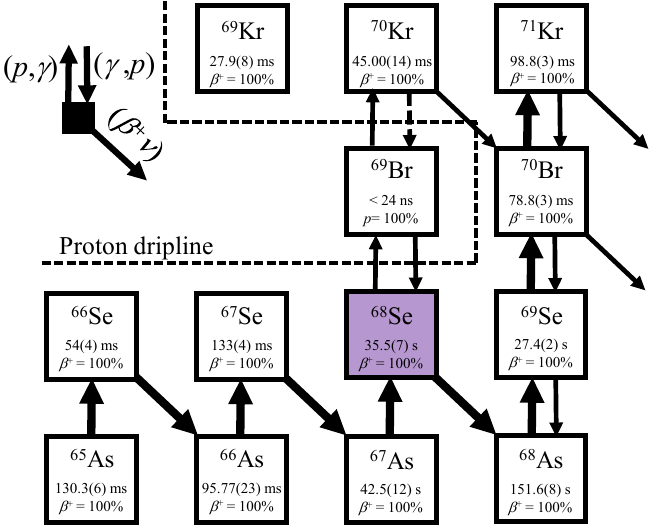}
	\caption{The $rp$-process path passing through the $^{68}$Se waiting point.}    
	\label{fig:path}
\end{figure}

The reaction flow passing $^{68}$Se is shown in Fig.~\ref{fig:path}. The $rp$-process primarily proceeds the pathway of $^{67}$Se($\beta^+$)$^{67}$As($p,\gamma$)$^{68}$Se($\beta^+$)$^{68}$As, competing with a 2$p$-capture process via $^{67}$Se($\beta^+$)$^{67}$As($p,\gamma$)$^{68}$Se($p,\gamma$)$^{69}$Br($p,\gamma$)$^{70}$Kr($\beta^+$)$^{70}$Br. To understand $rp$-process around the waiting point $^{68}$Se, several mass measurements of rare isotopes with $N\approx Z\approx 34$ have been performed with Penning trap mass spectrometry~\cite{Clark2004,Savory2009,Schury2007}. 
%In Ref.~\cite{Clark2004}, $^{68}$Se mass excess (ME) was determined to be $-54233(19)$~keV, and later was remeasured to be ME = $-54189.3(5)$~keV with a much improved precision. 
The mass excess (ME) of $^{68}$Se was determined to be  $-54189.3(5)$~keV. 
Combined with the predicted masses of $^{69}$Br and $^{70}$Kr based on isospin symmetry, effective lifetime of $^{68}$Se turned out to be about several tens seconds and it was concluded that the $^{68}$Se is a waiting point nuclide and presents a considerable delay in the $rp$-process. \citet{Schury2007} measured a series of $N > Z$ isotopes around $N= Z= 33$, yielding more precise and reliable masses for exotic $^{69}$Br and $^{70}$Kr through refined calculations of Coulomb displacement energy (CDE) using the improved masses of mirror nuclei. Consequently, the uncertainty in the effective lifetime of $^{68}$Se was reduced and a greater delay caused by this waiting point in the burst was shown. 

%The authors also stressed that the uncertainty in the effective lifetime of $^{68}$Se is dominated by uncertainty in the mass of $^{70}$Kr. 
%Following by 
%\xy{In} the work completed by \citet{Savory2009}, masses of $^{68}$Se with much more precise value was determined (ME = $-54189.3(5)$~keV). 
%Also determined in this work is the more precise mass value of $^{70}$Se, which is used to estimate the mass of $^{70}$Kr with the help of CDE~\cite{Brown2002}. 
%Then, a local network calculation was conducted using the updated mass data and reached a conclusion that $^{68}$Se is a strong waiting point \xy{under}
%for 
%all prevalent conditions.
It should be emphasized that the masses of relevant nuclides $^{69}$Br and $^{70}$Kr used in the aforementioned investigations were predicted values. The proton separation energy of the proton-unbound nucleus $^{69}$Br was first measured to be $S_p(^{69}{\rm Br}) = -785_{-40}^{+34}$ keV through complete kinematic measurement of its in-flight decay \cite{PhysRevLett.106.252503}. This key parameter $S_p(^{69}{\rm Br})$ was revised to be $-641(42)$ keV through a complementary approach detecting $\beta$-delayed protons of $^{69}$Kr \cite{DELSANTO2014453}. While both values agree reasonably with the CDE predictions of $-636(100)$ keV, the 140-keV increase in $S_p(^{69}{\rm Br})$ enhanced bypassing reaction flow branching ratio via $2p$ capture on $^{68}$Se inceasing from 0.3\% to 20\%. The mass of $^{69}$Br, which was indirectly obtained from its one-proton separation energy, was once strongly questioned since an anomalous odd-even staggering in CDE for the $T=1/2$ mirror nuclei was observed at $A = 69$ \cite{Tu_2014,PhysRevLett.110.172505}. This phase revision was rectified in a recent investigation of CDE with updated experimental masses \cite{PhysRevC.110.L021301,PhysRevC.111.014327}, validating the $S_p(^{69}{\rm Br})$ value obtained in the decay experiment.

Recently, the masses of a series of $T_z=-1$ $fp$-shell nuclides~\cite{Wang2023} including $^{70}$Kr were measured by using the newly established $B\rho$-defined isochronous mass spectrometry ($B\rho$-IMS)~\cite{Wang2022,Zhang2023,XZhou2024}. The mass excess of $^{70}$Kr is directly determined with the high-performance $B\rho$-IMS to be $-41320(140)$~keV, which is 220 keV more bound and more precise than AME2020 value ($\mathbf 41100\pm200$~keV).

In this {article}, we report the important role of the directly determined $^{70}$Kr mass in XRB models. It is the last piece of the puzzle of modelling the $^{68}$Se waiting point in $rp$-process. In particular the smaller $^{70}$Kr mass excess than AME2020 value unveils the non-negligible effect of the 2$p$-capture flow of $^{68}$Se. It reduces the effective lifetime $^{68}$Se, resulting in a 10\%-40\% change of the XRB luminosity and an enhancement of the final SnSbTe abundance in ashes due to a stronger $rp$-process heating.

\section{Calculation of effective half-life}\label{sec:calculation}
\begin{figure}[!htb]
	\centering
	\includegraphics[width=0.46\textwidth]{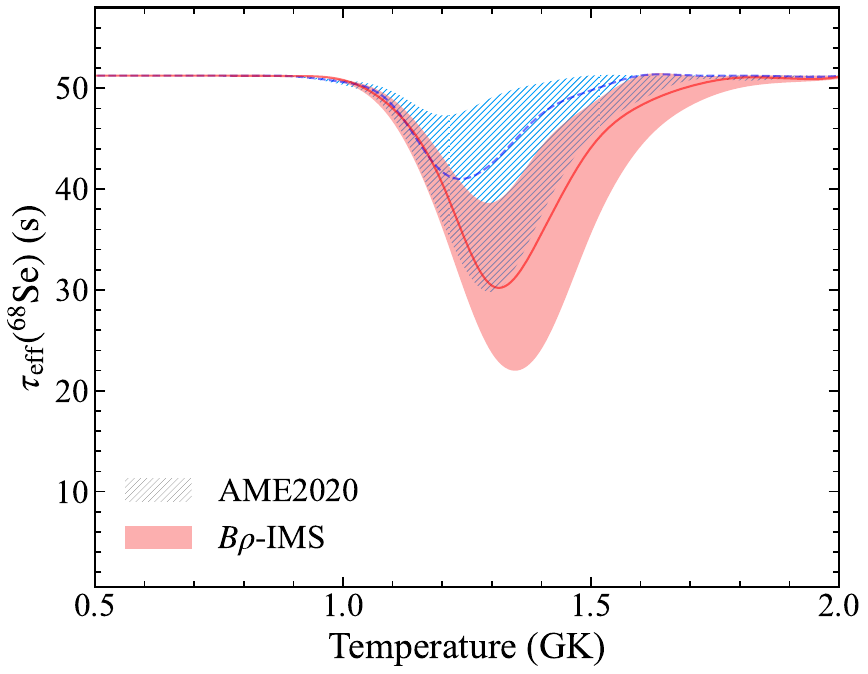}
	\caption{The effective lifetime of ${^{68} \rm Se}$ in the stellar environment as a function of temperature for typical $rp$-process conditions:$\rho=10^6 \rm\ g/cm^{3}$ and $Y_p=0.7$. The blue shaded region is the result using AME2020 mass and the corresponding 1$\sigma$ mass error. The red area is the range of lifetimes %within 
using the new mass measurement ($B\rho$-IMS) and the corresponding 1$\sigma$ uncertainties obtained in this work.}    
	\label{fig:t_eff}
\end{figure}

To determine the effective lifetime of $^{68}$Se, $\tau_{\rm eff}$($^{68}$Se), as a function of temperature, we performed a local small network calculation including proton capture reactions on $^{68}$As, $^{68,69}$Se,  $^{69}$Br, their corresponding photodisintegration reactions and $\beta$-decay of $^{68}$Se and $^{70}$Kr, similarly to the method in previous studies~\cite{Rodriguez2004,Schury2007}. We choose a typical density for XRB as $10^6 \rm\ g/cm^{3}$ and initial hydrogen abundance $Y_p=0.7$.  
The proton capture and the corresponding inverse reaction rates were taken from JINA Reaclib~\cite{Cyburt2010, REACLIB2010}, %except for $^{69}$Br($p,\gamma$)$^{70}$Kr which we calculate the rate based on the mass value of $^{70}$Kr in AME2020 and our new measurement by using the TALYS code~\cite{Koning2023}. 
while the reaction rate of $^{69}$Br($p,\gamma$)$^{70}$Kr was calculated by using the TALYS code~\cite{Koning2023} based on the mass value of $^{70}$Kr in AME2020 and $B\rho$-IMS measurement. 
The $^{70}$Kr($\gamma,p$)$^{69}$Br rate is calculated by using detailed balance. In Fig. \ref{fig:t_eff}, we show the temperature dependence of $\tau_{\rm eff}$($^{68}$Se), dashed and solid lines represent the results by using $m({^{70} \rm Kr})$ from AME2020 and the new measurement ($B\rho$-IMS), respectively. Red and blue regions are the 1$\sigma$ effective half-life variations originated from the corresponding mass uncertainties.

At low temperatures below 1 GK, proton captures are ineffective, so $\tau_{\rm eff}$($^{68}$Se) value is the $\beta$-decay lifetime of $^{68}$Se. For temperatures exceeding 1.5$\sim$2 GK, the photodisintegration starts to prevent proton capture flow. As a result, a $(p,\gamma)-(\gamma,p)$ equilibrium towards the waiting point forms,
%so the dominated flow is still $\beta$-decay.
so the dominated flow at $^{68}$Se is still $\beta$-decay. For intermediate temperatures, the proton capture plays important role in determining  $\tau_{\rm eff}$. The new measurement yielded smaller $^{70}$Kr mass value, resulting in a smaller forward reaction $Q$-value and thus smaller photodisintegration reaction rate. %harder to have photodisintegration reactions. 
Using the new experimental mass, the resultant effective lifetime is reduced by about $20\%$ compared with that using the AME2020 mass. 
%The obtained effective half-life for $^{68}$Se from both methods agree well with each other.

\section{XRB model calculation} 
\begin{figure}[!htb]
	\centering
	\includegraphics[width=0.48\textwidth]{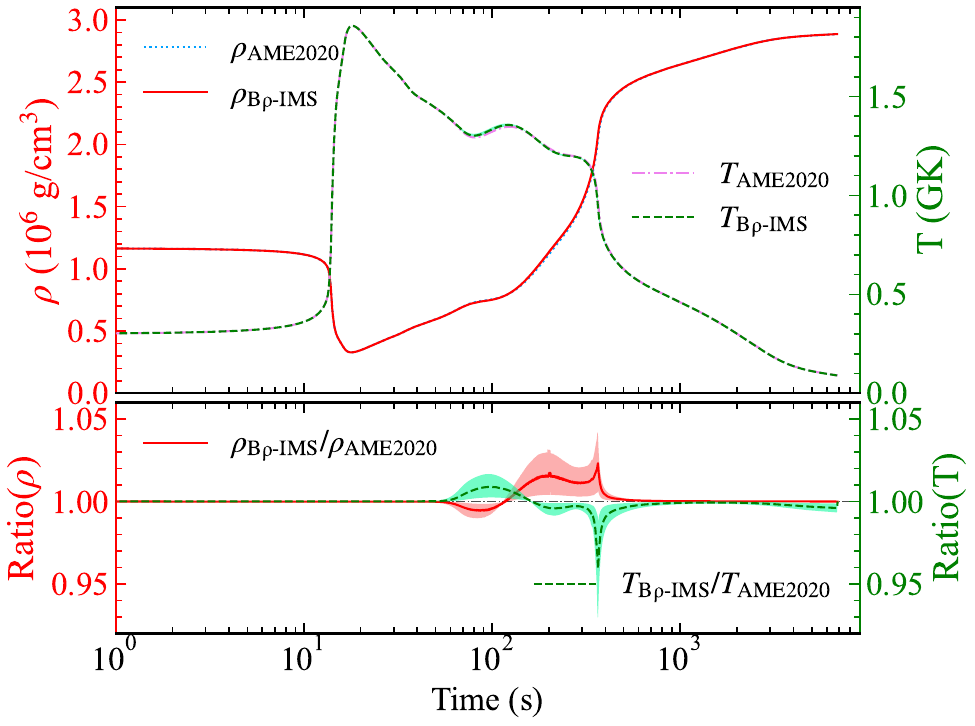}
	\caption{Top: The density and temperature trajectories from our one-zone XRB model. We take $^{70}$Kr mass mean value from AME2020 database and the $B\rho$-IMS, respectively. For the calculation, $P_{\rm ign}=10^{23.03}\  \rm dyn/cm^{2}$ and ratio of the H and He mass fraction was set to be $X/Y=3$. Bottom: The ratio of densities and temperatures between AME2020 and $B\rho$-IMS.}
	\label{fig:profile}
\end{figure}

%%%%%   Nuclide 897   %%%%%
\begin{table}[h]
\begin{center}
\caption{{Nuclides in nuclear reaction network implemented in our one-zone XRB code~\cite{2020PTEP.2020c3E02D}.}}
\label{table:897}
\renewcommand{\arraystretch}{1.1}
\begingroup
\small
\begin{tabular}{cccccccc}
\hline\hline
\multicolumn{2}{l}{Nuclide~~$A$} & \multicolumn{2}{l}{Nuclide~~~$A$} & \multicolumn{2}{l}{Nuclide~~~~$A$} & \multicolumn{2}{l}{Nuclide~~~~~$A$}\\
\hline
% 1
\multicolumn{1}{l}{$n$} & \multicolumn{1}{r|}{1} &
\multicolumn{1}{l}{S} & \multicolumn{1}{r|}{26 -- 37} & 
\multicolumn{1}{l}{Ge} & \multicolumn{1}{r|}{60 -- 75} &
\multicolumn{1}{l}{Cd} & \multicolumn{1}{r}{94 -- 113} \\  
% 2
\multicolumn{1}{l}{H} & \multicolumn{1}{r|}{1 -- 3} &
\multicolumn{1}{l}{Cl} & \multicolumn{1}{r|}{27 -- 38} &
\multicolumn{1}{l}{As} & \multicolumn{1}{r|}{61 -- 78} &
\multicolumn{1}{l}{In} & \multicolumn{1}{r}{95 -- 102} \\
% 3
\multicolumn{1}{l}{He} & \multicolumn{1}{r|}{3 -- 6} &
\multicolumn{1}{l}{Ar} & \multicolumn{1}{r|}{31 -- 41} &
\multicolumn{1}{l}{Se} & \multicolumn{1}{r|}{64 -- 81} &
\multicolumn{1}{l}{Sn} & \multicolumn{1}{r}{98 -- 112} \\
% 4
\multicolumn{1}{l}{Li} & \multicolumn{1}{r|}{6 -- 8} & 
\multicolumn{1}{l}{K} & \multicolumn{1}{r|}{32 -- 44} &
\multicolumn{1}{l}{Br} & \multicolumn{1}{r|}{66 -- 84} &
\multicolumn{1}{l}{Sb} & \multicolumn{1}{r}{99 -- 126} \\
% 5
\multicolumn{1}{l}{Be} & \multicolumn{1}{r|}{7 -- 11} & 
\multicolumn{1}{l}{Ca} & \multicolumn{1}{r|}{35 -- 45} &
\multicolumn{1}{l}{Kr} & \multicolumn{1}{r|}{69 -- 85} &
\multicolumn{1}{l}{Te} & \multicolumn{1}{r}{104 -- 127} \\
% 6
\multicolumn{1}{l}{B} & \multicolumn{1}{r|}{8 -- 13} &
\multicolumn{1}{l}{Sc} & \multicolumn{1}{r|}{36 -- 50} &
\multicolumn{1}{l}{Rb} & \multicolumn{1}{r|}{70 -- 88} &
\multicolumn{1}{l}{I} & \multicolumn{1}{r}{105 -- 129} \\
% 7
\multicolumn{1}{l}{C} & \multicolumn{1}{r|}{9 -- 15} &
\multicolumn{1}{l}{Ti} & \multicolumn{1}{r|}{39 -- 51} &
\multicolumn{1}{l}{Sr} & \multicolumn{1}{r|}{73 -- 89} &
\multicolumn{1}{l}{Xe} & \multicolumn{1}{r}{110 -- 130} \\
% 8
\multicolumn{1}{l}{N} & \multicolumn{1}{r|}{12 -- 18} &
\multicolumn{1}{l}{V} & \multicolumn{1}{r|}{41 -- 54} &
\multicolumn{1}{l}{Y} & \multicolumn{1}{r|}{74 -- 92} &
\multicolumn{1}{l}{Cs} & \multicolumn{1}{r}{111 -- 130} \\
% 9
\multicolumn{1}{l}{O} & \multicolumn{1}{r|}{13 -- 19} & 
\multicolumn{1}{l}{Cr} & \multicolumn{1}{r|}{42 -- 55} &
\multicolumn{1}{l}{Zr} & \multicolumn{1}{r|}{77 -- 93} &
\multicolumn{1}{l}{Ba} & \multicolumn{1}{r}{115 -- 130} \\
% 10
\multicolumn{1}{l}{F} & \multicolumn{1}{r|}{14 -- 22} &
\multicolumn{1}{l}{Mn} & \multicolumn{1}{r|}{43 -- 58} &
\multicolumn{1}{l}{Nb} & \multicolumn{1}{r|}{79 -- 97} &
\multicolumn{1}{l}{La} & \multicolumn{1}{r}{116 -- 130} \\
% 11
\multicolumn{1}{l}{Ne} & \multicolumn{1}{r|}{17 -- 23} & 
\multicolumn{1}{l}{Fe} & \multicolumn{1}{r|}{46 -- 59} &
\multicolumn{1}{l}{Mo} & \multicolumn{1}{r|}{82 -- 98}  &
\multicolumn{1}{l}{Ce} & \multicolumn{1}{r}{118 -- 130} \\ 
% 12
\multicolumn{1}{l}{Na} & \multicolumn{1}{r|}{18 -- 26} & 
\multicolumn{1}{l}{Co} & \multicolumn{1}{r|}{47 -- 62} &
\multicolumn{1}{l}{Tc} & \multicolumn{1}{r|}{83 -- 102} &
\multicolumn{1}{l}{Pr} & \multicolumn{1}{r}{119 -- 130} \\
% 13
\multicolumn{1}{l}{Mg} & \multicolumn{1}{r|}{19 -- 27} & 
\multicolumn{1}{l}{Ni} & \multicolumn{1}{r|}{49 -- 65} &
\multicolumn{1}{l}{Ru} & \multicolumn{1}{r|}{86 -- 103} &
\multicolumn{1}{l}{Nd} & \multicolumn{1}{r}{121 -- 130} \\
% 14
\multicolumn{1}{l}{Al} & \multicolumn{1}{r|}{22 -- 30} &  
\multicolumn{1}{l}{Cu} & \multicolumn{1}{r|}{51 -- 68} &
\multicolumn{1}{l}{Rh} & \multicolumn{1}{r|}{88 -- 106} &
\multicolumn{1}{l}{Pm} & \multicolumn{1}{r}{122 -- 130} \\
% 15
\multicolumn{1}{l}{Si} & \multicolumn{1}{r|}{23 -- 31} &  
\multicolumn{1}{l}{Zn} & \multicolumn{1}{r|}{55 -- 71} &
\multicolumn{1}{l}{Pd} & \multicolumn{1}{r|}{90 -- 109} &
\multicolumn{1}{l}{Sm} & \multicolumn{1}{r}{128 -- 130} \\
% 16
\multicolumn{1}{l}{P} & \multicolumn{1}{r|}{24 -- 34} &
\multicolumn{1}{l}{Ga} & \multicolumn{1}{r|}{56 -- 74} &  
\multicolumn{1}{l}{Ag} & \multicolumn{1}{r|}{92 -- 112}
 & & \\
%%%%%%%%%%%%%%%%%%%
\hline
\end{tabular}
\endgroup
\end{center}
\end{table}
%%%%%   Nuclide 897   %%%%%
To clarify the role of the new $^{70}$Kr mass value in the $rp$-process near $^{68}$Se, we further conducted one-zone XRB model calculations~\cite{Koike1999,2020PTEP.2020c3E02D}. 
Our one-zone code evolves thermodynamic values and each mass fraction in a single zone at a constant pressure $P=P_{\rm ign}$, neglecting temperature, density, and compositions gradients. {Our reaction network contains 897 nuclei as shown in TABLE~\ref{table:897}.} 
The metallicity of accreting matter was set to be $Z_{\rm CNO} = 10^{-3}$ and ratio of the H and He mass fraction was set as $X/Y=3$. The ignition pressure $P_{\rm ign}=10^{23.03}\ \rm dyn/cm^{2}$ remains constant during the flash with surface gravity $\log g_s = 14.38$. The SnSbTe cycle around $A\approx 100$ could build during the flash under such a parameter setting, which is consistent with the results of Ref.~\cite{Schatz2001}.
The density and temperature trajectories of this model are shown in Fig. \ref{fig:profile}. Blue dotted line and red solid line correspond to the density trajectories generated by using $^{70}$Kr mass from AME2020 and $B\rho$-IMS, respectively,
while violet dash dotted line and green dash line stand for temperature trajectories.
Deviations can be found around $t\approx50-450~{\rm s}$.

%\begin{figure}[!htb]
%	\centering
%	\includegraphics[width=0.46\textwidth]{flow.pdf}
%	\caption{The abundance flow calculated by using $^{70}$Kr mass in AME2020 database and in the new measurement using $B\rho$-IMS. We take $^{70}$Kr mass mean value for the calculation and $P_{\rm ign}=10^{23.03}\  \rm dyn/cm^{2}$ .}
%	\label{fig:flow}
%\end{figure}
%In Fig. \ref{fig:flow}, the abundance flow are presented for AME2020 and IMS cases. About 3\% abundance flow goes to $^{70}$Kr via 2$p$-capture compared with the $\beta$-decay of  $^{68}$Se in AME2020 case. The flow toward 2$p$-capture is enhanced to about 12\% when using our new measured $^{70}$Kr mass. 
%\yd{We present the abundance flow \AD{around ${}^{68}{\rm Se}$} in Fig.~\ref{fig:flow},} 
%The abundance flow surrounding $^{68}$Se are depicted 
%depicting for two scenarios: one using the $^{70}$Kr mass from AME2020 and another utilizing our $B\rho$-IMS experimental measurement. 
We find, at the ignition pressure of $P_{\rm ign}=10^{23.03}\  \rm dyn/cm^{2}$, 
%the ratio of the reaction flow into $^{70}$Kr via the 2$p$-capture channel to the reaction flow through the $^{68}$Se($\beta^+$) channel is about 3\% in the AME2020 case.
the reaction flow ratio of the 2$p$-capture on $^{68}$Se channel to the $^{68}$Se($\beta^+$) channel is about 3\% with $^{70}$Kr mass from AME2020.
%The ratio is increased to 12\% when the new experimental $^{70}$Kr mass is used. 
In contrast, adoption of the new experimental $^{70}$Kr mass elevates this ratio to 12\%.
This is because the $B\rho$-IMS experimental result has a higher proton separation energy $S_p$($^{70}$Kr) compared with AME2020, which enhances the flow towards $\beta$-decay of $^{70}$Kr due to lower $^{70}$Kr($\gamma,p$)$^{69}$Br rate.

%the $^{70}$Kr is more difficult to have $^{70}$Kr($\gamma,p$)$^{69}$Br reaction, enhance the flow towards $\beta$-decay of $^{70}$Kr to heavier nuclei.}

\begin{figure}[htb]
	\centering
	\includegraphics[width=0.46\textwidth]{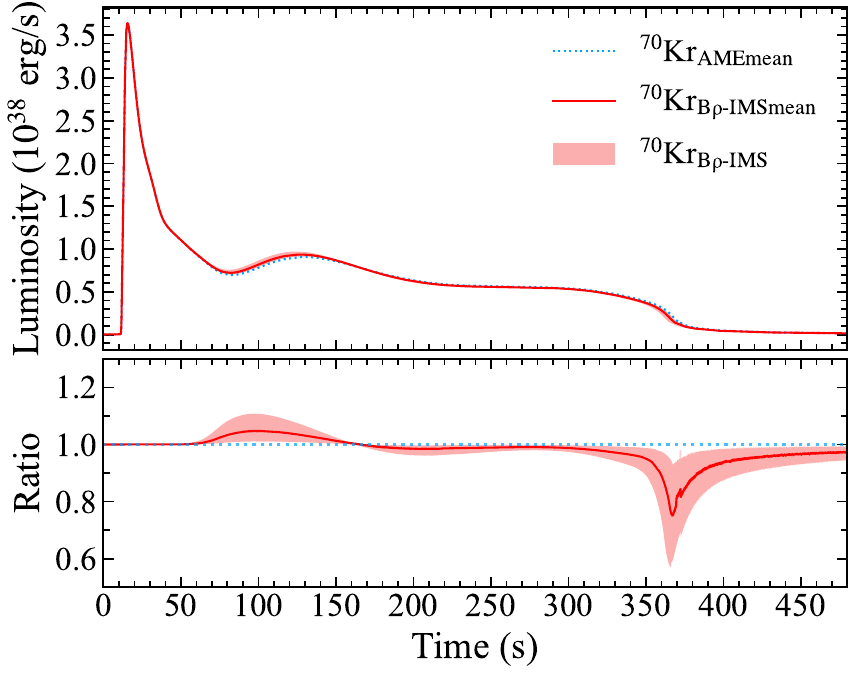}
	\caption{X-ray luminosity as a function of time using $^{70}$Kr mass from AME2020 (blue dotted line) and new mass (with 1$\sigma$ uncertainty) measured by $B\rho$-IMS (red solid line with shadow). The bottom panel shows the ratio of these luminosities.}
	\label{fig:lum}
\end{figure}

Del Santo et al. (2014) \cite{DELSANTO2014453} found that the impact of $S_p$($^{69}$Br) on the XRB light curve is highly correlated with $^{70}$Kr mass, and in particular for a high $^{70}$Kr-mass regions, they claimed that $S_p$($^{69}$Br) is unimportant due to a higher $S_p$($^{69}$Br) value, i.e., lower rate of $^{69}$Br($\gamma$,$p$)$^{68}$Se. In our result, the smaller $^{70}$Kr mass compared with AME2020 facilitates more $2p$ capture flow, such an enhanced $2p$ flow could change XRB light curves, as shown in Fig.~\ref{fig:lum}.

\begin{figure}[htb]
	\centering
	\includegraphics[width=0.46\textwidth]{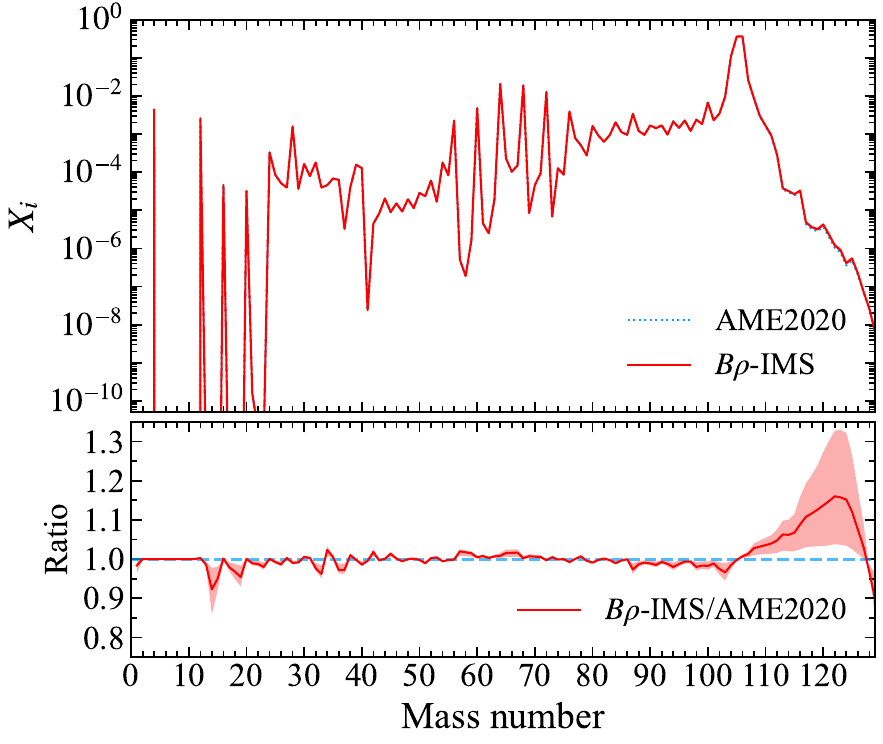}
	\caption{Comparison of the final {mass fraction} in ashes from AME2020 (blue dotted line) and new mass (with 1 $\sigma$ uncertainty) measured by $B\rho$-IMS (red solid line). The bottom panel shows the ratio of these final abundances.}
	\label{fig:abun}
\end{figure}

Figure.~\ref{fig:lum} shows calculated XRB light curves, comparing the results derived from two distinct $^{70}$Kr mass inputs: the AME2020 recommended mass (blue dotted line) and the newly measured mass with its 1$\sigma$ uncertainty obtained via the $B\rho$-IMS experiment (red solid line with shaded region). The discrepancy in the luminosity profiles highlights the critical influence of the $^{70}$Kr mass on the simulated XRBs. 
The light curve enlarged about  10\% at $t\approx100$ s due to a stronger $rp$-process heating from the higher flow from $^{70}$Kr towards heavier nuclei. This further leaves imprints in the tail structure around $t\approx370$ s as a faster consumption of hydrogen leading to a stronger SnSbTe cycle when the flow reaches nuclei with $A\approx100$. One can also see this effect in Fig. \ref{fig:abun}, where we compare the final {mass fraction} in the ashes for both $B\rho$-IMS experiment and AME2020 cases: The red line corresponds to the result obtained by using $B\rho$-IMS experiment data, which has an enhancement for SnSbTe elements in the final ashes compared with blue line.
{The impurity parameter $Q_{\rm imp}$ is often used to describe the degree of promiscuity. The $Q_{\rm imp}$ of the final products in our calculation is high to be $33.69^{+0.07}_{-0.04}$, which would show slow cooling in the accreting neutron stars after X-ray bursts, assuming that the $Q_{\rm imp}$ value is kept in such a cooling phase (e.g., \cite{Page:2012zt}).}

%around $N\approx Z\approx 34$ mass range
\section{Summary}

%In combination with the directly mass measurement of $^{70}$Kr using $B\rho$-IMS, the effective stellar half-life of the potential waiting point $^{68}$Se in the $rp$-process was re-estimated at both low and high stellar temperature of XRB model. 
%The rate of $^{69}{\rm Br}(p,\gamma)$ reaction was recalculated with TALYS code using $^{70}$Kr mass from both AME2020 and $B\rho$-IMS, two similar results were obtained and suggested that $^{68}$Se is
%not a strong waiting point in low temperature XRB models where the photodisintegration of $^{70}$Kr is negligible. 
%When the photodisintegration of $^{70}$Kr can not be neglected, the effective stellar half-life of $^{68}$Se was estimated in this work to be much longer than typical burst time scales, indicating that $^{68}$Se is a strong waiting point in the high temperature stellar conditions. 

%Although the $^{70}$Kr mass was directly measured by using $B\rho$-IMS, a relatively large uncertainty is persistent due to the its low production yield. 
%In order to more accurately quantify the extent to which $^{68}$Se delays the $rp$-process, a more precise mass measurement of $^{70}$Kr is desired.
%The theoretical mass values calculated from CDE and Guo $et~al$. methods are especially consistent with the directly measured value, indicating that the two methods can provide reliable mass predictions for the neutron-deficient nuclei not yet accessible to current experiments.
{Thanks to the advanced $B\rho$-IMS technique, the last piece of the puzzle of modelling the $^{68}$Se waiting point in $r$p-process was finally completed.
The experimental mass of neutron-deficient $^{70}$Kr is 220 keV lower than the AME2020 prediction, indicating a more tightly bound $^{70}$Kr and closely matching local mass predictions. With the newly measured $^{70}$Kr mass via $B\rho$-IMS, we re-evaluated the effective stellar half-life of the potential $r$p-process waiting point $^{68}$Se in XRB scenarios. The more-bound nature of $^{70}$Kr reduces the effective half-life of $^{68}$Se by a factor of two in the 1–2 GK temperature range under XRB conditions.}
%In summary, the experimental mass of neutron-deficient $^{70}$Kr for the first time is applied to XRB study. The more-bound nature of $^{70}$Kr reduces the effective half-life of $^{68}$Se thus a stronger $rp$-process heating could happen. The accelerated hydrogen consumption leads to a $10\% \sim 40\%$ luminosity difference alongside a strengthened SnSbTe cycle, resulting in a significant enhancement of SnSbTe abundances in the final ashes.

{A one-zone XRB model calculations demonstrate that the stronger rp-process heating driven by the updated $^{70}\text{Kr}$ mass leads to an $\sim 10\%$ luminosity increase at $t \sim 100 $~s. 
Furthermore, accelerated hydrogen consumption leaves imprints in the tail structure at $t \sim 370 $~s, 
alongside a strengthened SnSbTe cycle, resulting in a significant enhancement of SnSbTe abundances in the final ashes. The shaded uncertainty band in both Figs. \ref{fig:lum} and \ref{fig:abun} quantifies the propagated impact of the 1$\sigma$ experimental error in the  $^{70}$Kr mass, highlighting that the current substantial uncertainties in the mass measurement exert a pronounced influence on the SeBr cycle as well as the luminosity profile. This underscores the necessity of precise nuclear mass measurements for reliable astrophysical modeling. A more precise future mass measurement of $^{70}$Kr could help to clarify the role of other reaction rates in XRB model around $A\approx70$ mass range.}

\begin{acknowledgments}
	This work is financially supported in part by the National Key R$\&$D Program of China (Grants No. 2021YFA1601500, No. 2023YFA1606401, No. 2022YFA1602401), 
	the Natural Science Foundation for Young Scientists of Gansu Province, China (Grant No. 24JRRA036),
    the Youth Innovation Promotion Association of the Chinese Academy of Sciences (Grants No. 2022423, No. 2021419, No. 2019406), 
    and the NSFC (Grants No. 12135017, No. 12121005, No. 12322507, No. 11975280, No. 11905261, No. 12335009, No. 12435010), Regional Development Youth Program of the Chinese Academy of Sciences (No. [2023]15). Y. L is supported by Boya fellowship of Peking University and the China Postdoctoral Science Foundation under Grant Number 2025T180924. A.D. is supported by JSPS KAKENHI (Grant No. J25K17403).
\end{acknowledgments}


\bibliographystyle{apsrev4-2}
\bibliography{ref}

\end{document}